# Enhancing Spin Diffusion in GaAs Quantum Wells: The Role of Electron Density and Channel Width

B. W. Grobecker[1], A. V. Poshakinskiy[2], S. Anghel[1], T. Mano,[3] G. Yusa[4] and M. Betz[1]

[1]*Experimentelle Physik 2, Technische Universität Dortmund, Otto-Hahn-Straße 4a, D-44227 Dortmund, Germany*
[2]*ICFO-Institut de Ciencies Fotoniques, The Barcelona Institute of Science and Technology, 08860 Castelldefels, Spain*
[3]*National Institute for Materials Science, Tsukuba, Ibaraki 305-0047, Japan*
[4]*Department of Physics, Tohoku University, Sendai 980-8578, Japan*
E-mail address: markus.betz@tu-dortmund.de, sergiu.anghel@tu-dortmund.de

**Abstract:** This study explores the relationship between spin diffusion, spin lifetime, electron density and lateral spatial confinement in two-dimensional electron gases hosted in GaAs quantum wells. Using time-resolved magneto-optical Kerr effect microscopy, we analyze how Hall-bar channel width and back-gate voltage modulation influence spin dynamics. The results reveal that the spin diffusion coefficient increases with reduced channel widths, a trend further amplified at lower electron concentrations achieved via back-gate voltages, where it increases up to 150% for the narrowest channels. The developed theoretical model confirms the spatial inhomogeneities in the spin diffusion as arising from electron-density variations within the channels. The results underscore the importance of tuning electron density and spatial geometry to optimize spin transport and coherence, providing valuable design considerations for spintronic devices where efficient spin manipulation is crucial.

## I. INTRODUCTION

In low-dimensional non-centrosymmetric semiconductor structures, in particular in two-dimensional electron gases (2DEGs) within quantum wells (QWs), the spin-orbit (SO) interaction governs a wide range of spin-related phenomena[1-3]. Among these is the persistent spin helix (PSH) [4,5], a helical spin texture that emerges when the Rashba [6] and Dresselhaus [7] SO couplings are equal in magnitude. This condition restores SU(2) spin rotation symmetry [8], suppresses spin relaxation due to the Dyakonov–Perel mechanism, and creates robust spin coherence with a spatially striped spin texture [4]. The PSH state has been experimentally demonstrated through techniques like transient spin grating spectroscopy [5,9] and Kerr-rotation microscopy [4,10-28], showcasing its long-lived nature and potential for spintronic applications.

The PSH arises due to the unidirectional momentum-dependent effective magnetic field $B_{SO}(\boldsymbol{k})$, which protects the spin state from dephasing caused by spin-independent scattering. This protection, coupled with the ability to tune Rashba and Dresselhaus interactions through quantum well design, gate voltage modulation [20,29], or optical doping [19,21], allows for precise control of PSH dynamics. Studies have demonstrated that factors such as the spin diffusion coefficient [15], cubic Dresselhaus terms [30,31], and carrier heating [16] influence PSH lifetimes and diffusion properties. Adjustments in carrier density, excitation energy, and applied electric fields also impact the transition from ballistic to diffusive motion [11], modulating the spin transport efficiency. Recent advances link PSH dynamics to structured light and topological concepts [10] by using a spatially structured light such as vector vortex beam that enables spatially variant polarization states, offering applications in optical communication, metrology, and spintronic devices. Future devices could harness the synergy of long-lived PSH spin states and spatial spin modes for applications in information processing, quantum computing, and advanced optics. However, ensuring robust spin transport and coherence remains challenging, particularly due to quantum well design constraints that affect electron mobility, spin diffusion, and PSH lifetime ($\tau_s$).

A possible way to further increase the lifetime of spin textures is to consider quantum wells with lateral potential that confines electron gas to quasi-1D channels of a few-micron width. Such channels are still much wider than the electron-mean-free path and preserve the diffusive character of transport. However, they can be narrower than the spin-orbit precession length, leading to suppression of the Dyakonov–Perel mechanism spin relaxation mechanism [32-36]. It was theoretically predicted that in this regime the spin lifetime should increase with a power law $1/d^2$ where $d$ is the channel width [32]. The experimental studies also evidenced the spin lifetime enhancement [24,25,37], though limited by the effect of $k$-cubic spin-orbit interaction [24]. Importantly, in all studies the spin diffusion coefficient ($D_s$) has been assumed to be independent from the channel width.

The current paper seeks to advance knowledge in this area by investigating how channel width of the Hall-bar, combined with back-gate voltage modulation, affects the spin lifetime and spin diffusion coefficient. Using time-resolved magneto-optical Kerr microscopy, we find out that $D_s$ increases significantly as the width of the Hall-bar channels decreases. This increase is further enhanced when the density of the two-dimensional electron gas is decreased by the back-gate voltage. Since the spin-relaxation rate in Dyakonov—Perel mechanics is proportional to $D_s$, the increase of the latter in the narrow channels tries to speed up the spin relaxation, competing with the described effect of the slowing down of the spin relaxation.

## II. EXPERIMENTAL DETAILS

The sample under investigation is a modulation-doped (001)-oriented 15-nm GaAs quantum well, grown by molecular beam epitaxy and sandwiched between $Al_{0.33}Ga_{0.67}As$ barriers. The QW is patterned in a series of five channels with different widths as shown in Fig. 1(d). with AuGeNi ohmic contacts. The channels are oriented along $[1\bar{1}0]$ crystallographic direction, i.e. the direction of the spatial spin precession when the PSH conditions are met. Two Si $\delta$-doping layers are placed above the QW providing a resident electron concentration $n$ in the QW that can be modified by the back-gate voltage $U_{BG}$ [38]. Using magneto-transport measurements we very that both $n$ and electron mobility $\mu$ have a linear dependence on the back-gate voltage in the range of $1.5\,V < U_{BG} < -2.5\,V$. To create robust electron spins, the sample resides in a compact cold-finger cryostat ensuring a lattice temperature of 3.5 K for all performed measurements.

The time-resolved magneto-optical Kerr microscopy (TR-MOKE) measurements are performed using pulses with a temporal width of ~35 fs derived from a 60 MHz mode-locked Ti:Sapphire oscillator. Subsequently, they are split into pump and probe paths, which are spectrally tuned independently by grating-based pulse shapers [22]. The resulting pulses have a bandwidth of ~0.5 $nm$ and allow for a transform-limited temporal resolution of ~1 $ps$. The probe pulses are linearly polarized while the pump pulses are modulated between *left ($\sigma^+$)* and *right ($\sigma^-$)* circular polarization by an electro-optical modulator (EOM). Both probe and pump pulses are collinearly focused on the sample surface through a 50× microscope objective. The full width at half-maximum (FWHM) diameter of pump and probe pulses are $w_0 = 3 \pm 0.1\,\mu m$ and $1 \pm 0.1\,\mu m$ respectively. The reflected pump light is filtered out with a monochromator and the Kerr-rotation of the reflected probe pulse is measured using balanced photodiodes connected to a lock-in amplifier referenced to the EOM frequency. The delay time $t$ between the pump and probe pulses is adjusted by a mechanical delay stage with $t_{max} = 1.8$ ns. The spatial overlap of the pump with the fixed and centered probe is adjusted through a lateral translation of the input lens of a beam-expanding telescope in the pump path [39,40]. The pump and probe photon energies are chosen based on the spectral response of the 2DEG, see for example Ref [16].

All measurements are performed with the pump photon energy set to $E_p = 1.57$ eV, which is 40 meV above the bandgap energy (1.53 eV), and a peak power density of $I_p = 4.7$ MW/cm$^2$. Since the energy separation between the first and second electron levels in the QW is about 52 meV, the pump excites electrons to the first sublevel only. The probe photon energy is tuned to $E_{pr} = 1.53$ eV with a pulse peak irradiance of $I_{pr} = 2.36$ MW/cm$^2$.

## III. RESULTS AND DISCUSSION

Firstly, the sample is investigated at $U_{BG} = -1.55$ V, corresponding to an electron concentration of $n = 3.25 \times 10^{11}$cm$^{-2}$. This aimed to ensure a longest possible spin lifetime under the given experimental conditions (note that this voltage does not correspond to the PSH condition, as was shown in [15]). Figures 1(a), (b) and (c) show exemplary spatial evolutions of the measured spin polarization $S_z$ within three selected Hall-bar channels of different widths. Here, the time delay between the initial pump pulse and the scanning probe pulse was fixed to $t = 570$ ps. This makes sure that the helical spin pattern has enough time to establish itself. The dimension of each respective channel is highlighted by a dotted line. The existence of some signal outside of those borders corresponds to the relatively big pump diameter of $3 \pm 0.1$ μm. Especially for the narrowest channels, this lies within the order of magnitude of channel width $d$. Therefore, by using a probe-stationary approach, spin polarization is induced by the remaining overlap of pump spot and embedded channel, although the center of the pump pulse is already off.

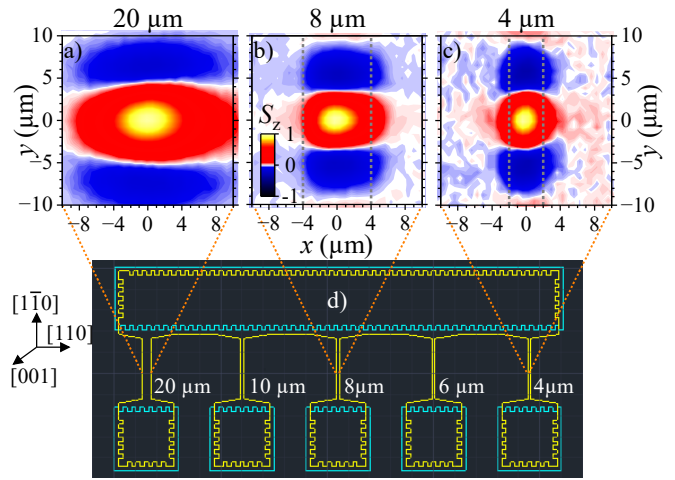

Figure 1: Channel mask with five different channel widths of the investigated GaAs QW sample (the yellow and cyan lines correspond to mesa and ohmic contacts, respectively) (d) and exemplary 2D spatial maps of the spin polarization $S_z$ at a delay time of $t = 570$ ps for the channels widths of (a) 20 μm (b) 8 μm and (c) 4 μm respectively.

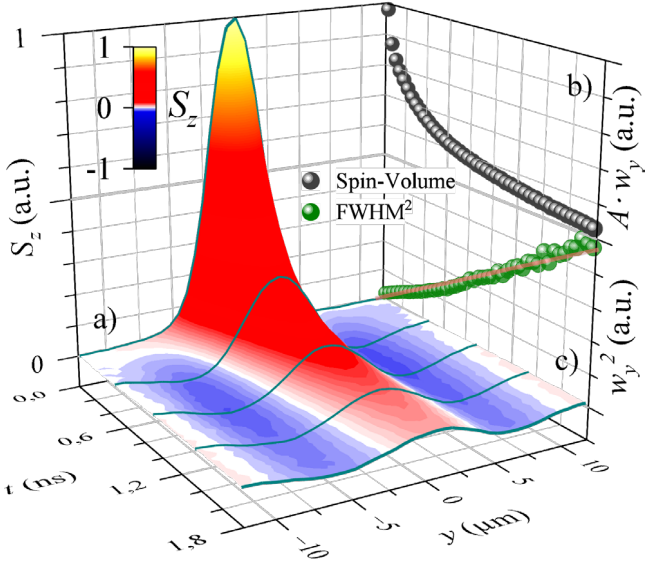

Figure 2: (a) Spatio-temporal mapping of the induced spin polarization distribution $S_z(t, y)$ of the 8 µm channel. For better visualization five arbitrary spatial scans are highlighted. For any given delay time $t$ the spatial dependence gets fitted to Eq. (1). Two of the resulting fit parameters $(A, w_y)$ are used to display (b) the spin volume $A \cdot w_y$ and (c) the square of FWHM. The z axis is restricted to arbitrary units and only the time axis is shared by all plots throughout the figure.

To investigate the interplay of the channel dimensions and the electron density in manners of spin dynamics, it is essential to map the latter in both space and time. Given spatial constraints imposed by the channel widths, the spatial resolution is limited to the $y$ axis (1D case), which aligns along the crystallographic axis [1$\bar{1}$0]. Figure 2(a) showcases an exemplary spatiotemporal measurement for the 8 µm channel. The broadening and decay of the initial Gaussian shaped spin polarization $S_z(t, y)$ is caused by diffusion and relaxation mechanisms respectively. To visualize the spin distribution amplitude and broadening in space and time, five exemplary spatial cuts are shown superimposed with the original signal. At any given delay time $t$ the spatial spin distribution is fitted with the phenomenological equation

$$S_z(y) = A \cdot e^{-\frac{4\ln(2)\,(y-y_G)^2}{w_y^2}} \cos\left(\frac{2\pi(y-y_c)}{\lambda_{SO}}\right) \quad (1),$$

where $A(t)$ is the amplitude of the spin polarization, $w_y(t)$ is the FWHM of the Gaussian envelope and $\lambda_{SO}(t) = \lambda_0 w_y(t)^2/(w_y(t)^2 - w_0^2)$ is the momentary spin precession length, whereas $\lambda_{0,y} = \pi\hbar^2/(m^*|\alpha + \beta|)$ is the precession length of the long-lived spin-helix. Due to the significant spatial restrictions imposed by the investigated Hall-bar channels, especially when it comes to the narrow channels, we assume that spatial dynamics mostly occur only in one dimension. Therefore, the temporal evolution of the total number of spins can be quantified by $A \cdot w_y(t)$. This expression accounts for all initially excited spins, not only those that are oriented along $z$ at the moment of detection and contribute to measured $S_z$ but also those spins, which, due to the spatial precession, lie in the QW plane. This total spin-volume, whose time dependence is shown in Fig. 2(b), allows us to retrieve the spin lifetime $\tau_s$ by fitting it with a single exponential decay. Furthermore, the time-dependent square of the Gaussian envelope FWHM, illustrated in Fig. 2(c), provides access to the spin diffusion coefficient $D_s$ via a linear regression of

$$w_y^2(t) = w_0^2 + 16\ln(2)D_s t \quad (2).$$

The same procedure of analyzing the experimental data is applied to all the subsequent measurements done for each channel seen in Fig. 1(d).

This paper aims to investigate the dependence of the spin diffusion coefficient, spin lifetime, and electron concentration $n$ on the channel width $d$ and various back-gate voltages. Specifically, it focuses on the relative changes in these parameters compared to their behavior in the absence of spatial restrictions imposed by channel width. To this end, measurements were conducted for different channel widths while varying the back-gate voltage $U_{BG}$ between $+0.5$ V and $-2.0$ V. Figure 3 presents the experimental results as ratios relative to the corresponding parameter values obtained for $d = 20$ µm (a channel width that does not impose significant lateral confinement). These reference values are denoted by a subscripted zero ($D_0, \tau_0$). The first striking result is that $D_s/D_0$ ratios increase drastically with limiting the spatial dynamics in one dimension: the smaller $d$ the higher the $D_s$. Additionally, when different $U_{BG}$ were applied, it can be observed that this effect is further enhanced by decreasing the $U_{BG}$, i.e., by reducing $n$. Hence, the range of amplification varies from only ~1/4 (for a $U_{BG} = 0.5$ V) up to ~3/2 ($U_{BG} = -2.0$ V) at the narrowest channel width of 4 µm. It gives the impression that the narrower the channel, the greater the influence of the electron concentration on the acceleration of diffusion in one dimension. In Fig. 3(b) the ratios of $\tau_s/\tau_0$ are visualized. Here the curve of $U_{BG} = -1.55$ V stands out as it shows a rather voltage-independent ratio of 1.0 with a small deviation at $d = 10$ µm. This can be attributed to the maximum spin lifetime, which was observed at this specific voltage during the characterization of the sample. The remaining voltages tend to allow a small rise in the spin lifetime by reducing the channel width. The parameters obtained for high electron concentrations ($U_{BG} = 0.5$ V) show large errors, since the fitting of the spin volume obtained here would provide better results when fitted by a twofold exponential decay. However, to ensure a standardized procedure for a better comparison, all the fittings were done using a single exponential decay.

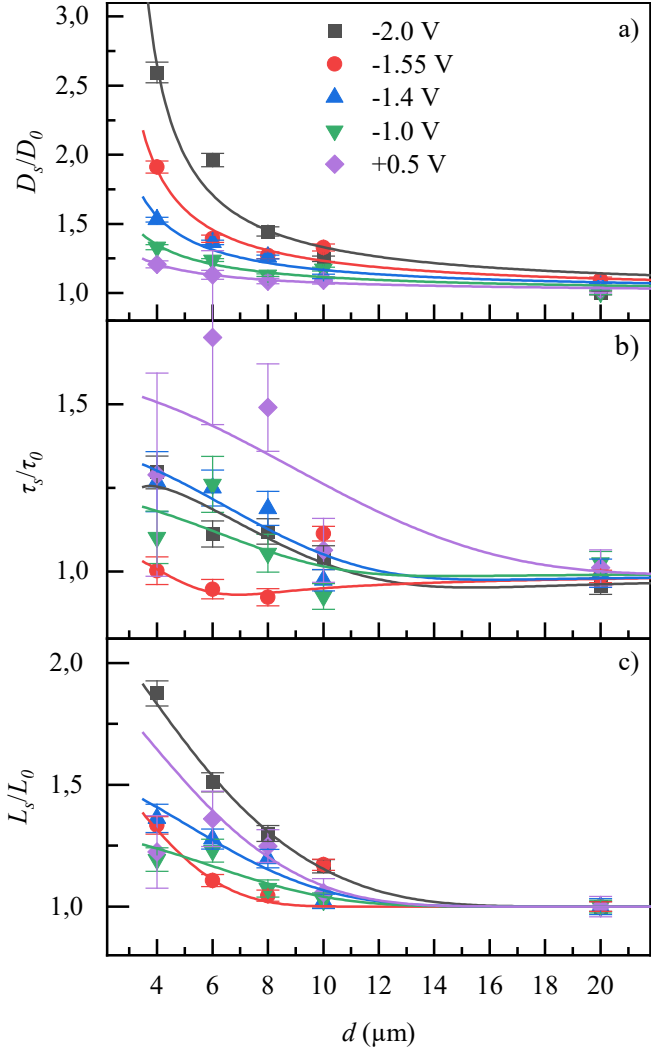

Figure 3: Results from analyzing the interplay of channel width $d$ and back-gate voltage $U_{BG}$ in manners of spin dynamic parameters (a) spin diffusion coefficient ratio $D_s/D_0$, (b) spin lifetime ratio $\tau_s/\tau_0$ and (c) spin diffusion length ratio $L_s/L_0$ ($L = \sqrt{D\tau_s}$) where $D_0, L_0$ and $\tau_0$ are the corresponding parameters obtained for the widest channel $d = 20$ μm.

Finally, the ratio of spin diffusion length $L_s/L_0$ (where $L = \sqrt{D\tau_s}$) is displayed in Fig. 3(c). For low $n$ ($U_{BG} = -2.0$ V) and small $d$ (4 μm) the ratio gets almost doubled, in comparison to having no lateral restriction. Due to the rather stable lifetime at $U_{BG} = -1.55$ V, here the extension of $L_s$ is not as pronounced. Still, it reaches ~1/3, due to the sudden increase of $D_s$ at $d = 4$ μm. The ratios for remaining voltages lie in between, although the spike for the smallest channel is suppressed. It emerges as a rather steady increase by decreasing $d$.

To explain the above seen relation between $U_{BG}$, $D_s$, and channel width $d$, we develop a theoretical description of spin diffusion of the 2D electrons inside the channel $-d/2 < x < d/2$, formed by the external potential $U(x)$ and described by a spatially inhomogeneous spin diffusion coefficient $D_s(x)$, see Fig. 4(a). We suppose that the channel has a symmetric profile, so $U(x) = U(-x)$ and $D_s(x) = D_s(-x)$.

The diffusion equation for the spin distribution function $S(x, y, t)$, accounting for the spin-orbit interaction, reads

$$\frac{\partial \mathbf{S}}{\partial t} = \sum_{\alpha=x,y}\left(\mathbf{\Lambda}_\alpha - \frac{\partial}{\partial r_\alpha}\right)D_s\left(\mathbf{\Lambda}_\alpha - \frac{\partial}{\partial r_\alpha} - \frac{1}{T}\frac{\partial U}{\partial r_\alpha}\right)\mathbf{S}, \quad (3)$$

where tensor $\mathbf{\Lambda}_\alpha$ describes spin rotation in the spin orbit field, $\mathbf{\Lambda}_{x,y}\mathbf{S} = 2m^*(\beta \mp \alpha)\mathbf{e}_{y,x} \times \mathbf{S}/\hbar^2$. We also suppose that the electron gas is non-degenerate and described by the temperature $T$ (measured in the energy units). In case of the degenerate electron gas, $T$ should be replaced by the Fermi energy $E_F(x)$. The spin diffusion equation (3) should be accompanied by the boundary condition,

$$\left(\mathbf{\Lambda}_x - \frac{\partial}{\partial x} - \frac{1}{T}\frac{\partial U}{\partial x}\right)\mathbf{S}(r)\bigg|_{x=\pm d/2} = 0, \quad (4)$$

which reflects the conservation of the spin at the channel edge. We search for the eigen solutions of the diffusion equation that have the form

$$\mathbf{S}_k(x, y, t) = e^{-U(x)/T + iky + \Lambda_x x - \Gamma_k t} \mathbf{P}_k(x). \quad (5)$$

Here, the spin polarization degree $\mathbf{P}_k(x)$ should be obtained by solving the Sturm-Liouville problem

$$-e^{-U(x)/T}\frac{d}{dx}e^{U(x)/T}D_s(x)\frac{d\mathbf{P}_k(x)}{dx}$$
$$+ \left(k + e^{-\Lambda_x x}i\mathbf{\Lambda}_y e^{\Lambda_x x}\right)^2 D_s(x)\mathbf{P}_k(x) = \Gamma_k \mathbf{P}_k(x). \quad (6)$$

Since there is no analytical solution for this equation in the general case, we use the perturbation theory, assuming $\|\mathbf{\Lambda}_x w\| \ll 1$ and following the approach of Ref. [32]. First, we describe unperturbed solutions, i.e., for $\mathbf{\Lambda}_x = 0$ which corresponds to the PSH regime. The solutions with the lowest decay rate have homogeneous distribution of spin polarization across the channel, $\mathbf{P}_k(x) = \mathbf{p}^{(\sigma)}$, where $\mathbf{p}^{(\sigma)}$ is an eigenvector of $\mathbf{\Lambda}_y$ and satisfies $\mathbf{\Lambda}_y \mathbf{p}^{(\sigma)} = ik_0 \mathbf{p}^{(\sigma)}$ with $k_0 = \sigma m(\alpha + \beta)$, $\sigma = 0, \pm 1$, and $|\mathbf{p}^{(\sigma)}|^2 = 1$. The optically accessible modes, which have nonzero spin component along the $z$ axis, correspond to $\sigma = \pm 1$. Note that for $k = k_0$, the unperturbed decay rate of the modes vanishes $\Gamma_{k_0} = 0$, which is the signature of the PSH state. There are other solutions of the unperturbed problem, which are described by $\mathbf{P}_k(x)$ with a certain number of zeros and fast decay rates $\Gamma_k \sim D_s/d^2$.

To consider the deviation from the PSH regime, we rewrite Eq. (6) as

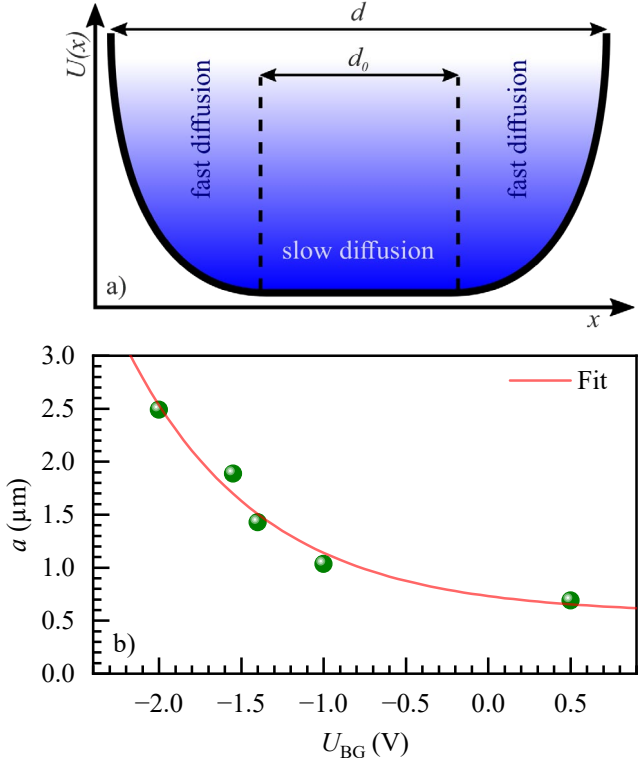

Figure 4: (a) Spatial profile of the external potential $U(x)$, which induces a spatially inhomogeneous spin diffusion coefficient $D_s(x)$. (b) Extracted fit parameter $a$ of Eq. (15) in dependence on the back-gate voltage $U_{BG}$.

$$-e^{-U(x)/T}\frac{d}{dx}e^{U(x)/T}D_s(x)\frac{d\boldsymbol{P}_k(x)}{dx}$$
$$+\left[(k_0+i\boldsymbol{\Lambda}_y)^2+\boldsymbol{V}(x)\right]D_s(x)\boldsymbol{P}_k(x)$$
$$=\Gamma\boldsymbol{P}_k(x), \qquad (7)$$

where
$$\boldsymbol{V}(x) = \{(k_0+i\boldsymbol{\Lambda}_y),(k'+e^{-\Lambda_x x}i\boldsymbol{\Lambda}_y e^{\Lambda_x x}-i\boldsymbol{\Lambda}_y)\}$$
$$+\left(k'+e^{-\Lambda_x x}i\boldsymbol{\Lambda}_y e^{\Lambda_x x}-i\boldsymbol{\Lambda}_y\right)^2, (8)$$

$k'=k-k_0$, and $\{A,B\}=AB+BA$. We expand $\boldsymbol{V}(x)$ up to terms quadratic in $k'$ and $\Lambda_x x$ and get

$$\boldsymbol{V}(x) = \left(k'-ix[\boldsymbol{\Lambda}_x,\boldsymbol{\Lambda}_y]\right)^2$$
$$+\left\{(k_0+i\boldsymbol{\Lambda}_y),\left(k'-ix[\boldsymbol{\Lambda}_x,\boldsymbol{\Lambda}_y]\right.\right.$$
$$\left.\left.-\frac{ix^2}{2}[\boldsymbol{\Lambda}_x,[\boldsymbol{\Lambda}_x,\boldsymbol{\Lambda}_y]]\right)\right\}. \qquad (9)$$

Next, we determine the effect of the perturbation $\boldsymbol{V}(x)$ on the unperturbed solution $\boldsymbol{p}^{(\sigma)}$. Note that for the considered mode $(k_0+i\boldsymbol{\Lambda}_y)\boldsymbol{p}^{(\sigma)}=0$, so only the first term in Eq. (9) contributes. The parity of the problem also constrains the mixing of the modes. Up to the order $\propto d^2$, the mixing of the $\boldsymbol{p}^{(\sigma)}$ with different $\sigma$ does not occur. The mixing of the $\boldsymbol{p}^{(\sigma)}$ mode with the fast-decaying modes of the odd parity leads to the contributions to the decay rate of the former of the order $\propto d^4$. Therefore, in the order $\propto d^2$, the decay of the eigen solution $\boldsymbol{p}^{(\sigma)}$ is determined by averaging the first line of Eq. (9), which yields $\Gamma_{k_0+k'}=\overline{\Gamma}+\overline{D}_s k'^2$, with the effective spin diffusion coefficient

$$\overline{D}_s = \frac{\int D_s(x)e^{-U(x)/T}dx}{\int e^{-U(x)/T}dx} \qquad (10)$$

and the spin relaxation rate
$$\overline{\Gamma} = \frac{16m^{*4}(1-\sigma^2/2)(\alpha^2-\beta^2)^2}{\hbar^8}$$
$$\times \frac{\int x^2 D_s(x)e^{-U(x)/T}dx}{\int e^{-U(x)/T}dx}. \qquad (11)$$

Finally, we suppose the spin diffusion coefficient is limited by electron-electron collisions, thus inversely proportional to the electron density $D_s(x)\propto e^{U(x)/T}$ [15]. For the optically accessible modes with $\sigma=\pm 1$, the above expressions for $\overline{D}_s$ and $\overline{\Gamma}$ are then simplified to

$$\overline{D}_s = D_s(0)\frac{d}{\overline{d}}, (12)$$
$$\overline{\Gamma} = \overline{D}_s\frac{2m^{*4}(\alpha^2-\beta^2)^2 d^2}{3\hbar^8}, (13)$$

where $d$ is the channel width, $\overline{d}=\int e^{-U(x)/T}dx$, and $U(0)=0$ is assumed.

The simplest model of the $U(x)$ profile is illustrated in Fig. 4(a). In this model, the potential is assumed to be flat ($U(x)=0$) in the central region of the channel, which has a width $d_0$, while it increases near the channel edges. As a result, the central region exhibits high electron density, leading to slower spin diffusion. In contrast, the lower electron density near the channel edges causes faster spin diffusion in those areas. $\overline{d}$ can be rewritten such as $\overline{d}=d-a$, where

$$a = \int (1-e^{-U(x)/T})\,dx \qquad (14)$$

is determined by the channel border only and assumed to not depend on the channel width $d$. Finally, we obtain the diffusion coefficient according to

$$\overline{D}_s = \frac{D_s(0)\,d}{d-a} \qquad (15)$$

which is valid for $d>d_0$.

We use Eq. (15) to fit the experimental curves in Fig. 3(a) and extract the fit parameter $a$ for different back gate voltages. The dependence of $a$ on $U_{BG}$ is shown in Fig. 4(b). With the increase of $U_{BG}$, the density of electrons grows, and they start to screen the channel profile potential $U(x)$, leading to the decrease of $a$, according to Eq. (14).

The change of the spin lifetime $\tau_s=1/\overline{\Gamma}$ with the channel width $d$, shown in Fig. 3(b), shall be explained by the competition of the factor $d^2$ in Eq. (13) and the decrease of $\overline{D}_s$. To eliminate the effect of $\overline{D}_s$, we plot in Fig. 3(c) the spin diffusion length $L_s=(\overline{D}_s/\overline{\Gamma})^{1/2}$ which, according to Eq. (13), should be described by

$$L_s = \frac{\sqrt{3/2}\,\hbar^4}{m^{*2}|\alpha^2 - \beta^2|d}. \qquad (16)$$

However, note that while the behavior of the diffusion coefficient $\overline{D}_s$ is well described by Eq. (12) in a wide range of channel width $d$, the applicability of Eqs. (13) and (16) is limited to the small value of $d$ only. Accordingly, only the initial part of the experimental data in Fig. 3(c) follows the $1/d$ trend, while for larger $d$ the spin diffusion length tends to a constant value, which corresponds to the 2D diffusion in the absence of spatial restrictions imposed by the channel widths.

## IV. CONCLUSIONS

This study offers a detailed examination of spin dynamics in quasi-1D channels within GaAs quantum wells, emphasizing the relationship between spin diffusion coefficient, spin lifetime, and electron density under varying geometrical and electronic conditions. Using TR-MOKE microscopy, the findings highlight the significant impact of lateral confinement and back-gate voltage modulation on spin transport. Specifically, the spin diffusion coefficient increases notably with narrower Hall-bar channel widths, reaching up to a 150% enhancement in the narrowest channels, an effect further amplified at reduced electron concentrations controlled via back-gate voltages.

A theoretical framework corroborates these experimental observations, demonstrating how spatial variations in electron density influence spin diffusion and relaxation. The model sugests two distinct regions within the channels: a central region, where electron density is high and spin diffusion is slow, and areas near the channel edges, where lower electron density results in faster spin diffusion. These spatially varying densities significantly shape spin transport behavior.

The findings underscore the potential for optimizing spin coherence and transport efficiency through careful tuning of electron density and spatially constrained geometries. This has profound implications for advancing spintronic devices. Future research could focus on refining these strategies, including innovations in material design and external modulation techniques, to harness these insights for real-world applications in spin-based technology.

## AKNOWLEDGMENTS

The authors are grateful to T. Takayanagi for the fruitful discussions and Y. Takahashi for the help with the sample preparation. This work is supported by and a Grant-in-Aid for Scientific Research (Grants Nos. 19H05603, 21H05182, 21H05188, and 24H00399) from the Ministry of Education, Culture, Sports, Science, and Technology (MEXT), Japan.